\documentclass{aa}
\usepackage{amsmath}
\usepackage{txfonts}
\usepackage{graphicx}
\usepackage{natbib}
\bibpunct{(}{)}{;}{a}{}{,}
\usepackage{array}
\usepackage{subfigure}
\usepackage{float}
\usepackage{txfonts,rotate,psfig,subfigure,supertabular,lscape,times}
\usepackage{epsfig}
\usepackage{aalongtable}

\setlongtables

\begin{document}

\title{High frequency VLBI observations of the scatter broadened quasar B\,2005+403}
\author{K.\,\'E. Gab\'anyi\inst{1} \and T.\,P. Krichbaum\inst{1} \and S. Britzen\inst{1} \and U. Bach\inst{1,2} \and E. Ros\inst{1} \and A. Witzel\inst{1} \and J.\,A. Zensus\inst{1}}

\institute{Max-Planck-Institut f\"ur Radioastronomie (MPIfR), Auf dem H\"ugel 69,
Bonn, 53121 Germany \and Istituto Nazionale Di Astrofisica (INAF), Osservatorio Astronomico di Torino, Via Osservatorio 20, 10025 Pino Torinese, Italy}
\offprints{K.\,\'E. Gab\'anyi, \email{gabanyik@mpifr-bonn.mpg.de}}
\date{Received / Accepted }

\abstract{

The quasar B\,2005+403 located behind the Cygnus region in our Galaxy, is
a suitable object for studying the interplay between propagation effects, which are
extrinsic to the source and source intrinsic variability. On the basis of VLBI
experiments performed at 1.6, 5, 8, 15, 22, and 43\,GHz during the years 1992 -- 2003 and parallel
multi-frequency monitoring of the total flux density (between 5 -- 37\,GHz), we investigated the variability
of total flux density and source structure. Below 8\,GHz, the point-like VLBI source is
affected by scatter-broadening of the turbulent interstellar medium (ISM), which is located
along the line of sight and likely associated with the Cygnus region. 
We present and discuss the measured frequency dependence of the source size, which shows 
a power-law with slope of $-1.91 \pm 0.05$. From the measured scattering angle at 1\,GHz of $\theta = 77.1 \pm 4.0$\,mas
a scattering measure of SM= $0.43 \pm 0.04 {\rm\,m}^{-20/3}$\,kpc is derived, consistent with the general properties
of the ISM in this direction. The decreasing effect of angular broadening towards higher frequencies
allows to study the internal structure of the source. Above 8\,GHz new 
VLBI observations reveal a one-sided slightly south-bending core-jet structure, 
with stationary and apparent superluminally moving jet components.
The observed velocities range from $6.3 - 16.8$\,c. The jet components move on non-ballistic trajectories.
In AGN, total flux density variations are often related to the emergence of new VLBI components.
However, during almost eleven years no new component was ejected in B\,2005+403.
A striking feature in the flux density variability is a trough observed at 5 -- 37\,GHz and between 1996 and 2001.
This trough is more pronounced at higher frequencies, where it shows a prominent flux density decrease of 
$\sim 50 - 60$\,\% on a time scale of $2-3$\,yrs. The trough can be explained as a blending
effect of decreasing and increasing jet component fluxes. Dense in time sampled flux density
monitoring observations with the 100\,m Effelsberg telescope reval intra-day variability (IDV) at 1.6\,GHz with 
a modulation index of $m=1.0$\,\%. At 5\,GHz less pronounced variations are seen ($m=0.5$\,\%).
This and a relatively short variability time scale of $\sim 0.1$\,days imply a second, less dense or turbulent scattering
screen at nearby (few to hundred parsec) distance.  

}

%\authorrunning{K.\,\'E. Gab\'anyi et al.} Nem tudom, hogy kell-e vagy nem, a referee pedanyba kellett, mert kerte
%\titlerunning{High frequency VLBI observations of the quasar B\,2005+403}

\maketitle

\section{Introduction}

The propagation of radio waves through the ionized interstellar medium causes
several effects, such as Faraday rotation and depolarization of polarized emission,
dispersion of pulsar signals, scatter broadening of compact radio sources and intensity
fluctuations caused by diffractive (DISS) and refractive (RISS) interstellar scintillation (ISS).
The effect of scatter broadening plays an important role in the understanding of the compact
object located in the center of our galaxy \object{Sgr\,A*} \citep[e.g.][]{galcenter2,galcenter3,galcenter}.
It is however also important for other galactic regions on the sky, e.g.
for the Cygnus region and the quasar B\,2005+403 located behind.
The detailed study of scatter broadening in compact radio sources could lead
to a better understanding of the interstellar medium. It also bears the potential to reveal fine details
of the intrinsic source structure, particularly at longer wavelengths, where direct imaging often
cannot provide sufficiently high angular resolution.

The other prominent propagation effect addressed in this paper is the so called
Intra-Day Variability \citep[IDV,][]{idv_discovery,idvh}, and
the question if and how it is related to the interstellar scintillation.
Recent IDV-surveys show that a large fraction (up to 30\,\%) of all compact
flat spectrum radio sources show this effect, which is characterized by
variability amplitudes of up to $20-30$\,\% and variability timescales
ranging from less than one hour to several days \citep[e.g.][]{reduc_idv,idvh,southern,lovell}.
If interpreted via source intrinsic incoherent emission processes, such short variability timescales
imply -- via the light travel time argument --  apparent brightness temperatures of
$T_\text{B} = 10^{16...19}$\,K, far in excess of the inverse-Compton limit \citep{compton}.
With the assumption of relativistic Doppler-boosting the brightness temperatures can be reduced,
which however, would imply uncomfortably large Doppler boosting factors of
$\delta \simeq 20 - 200$.

Alternately, the IDV in the radio-bands is also explained
extrinsically via scintillation of radio waves in the turbulent interstellar medium (ISM)
of our Galaxy \citep[e.g.][]{rick_uj,dt}. The main problem in this interpretation is
that it lacks the explanation of observed radio-optical broad-band correlations in at least some sources
(\cite{corr_var}, see also \cite{ww}). The recent detection of diffractive ISS in the source
J\,1819+3845 leads to micro-arcsecond source sizes and brightness temperatures of $\sim 10^{14}$\,K,
which again requires Doppler boosting factors of $\delta \simeq 100$ \citep{diff_1819}. 
It is therefore likely that the IDV phenomenon invokes both, a combination of
source intrinsic {\it and} extrinsic effects \citep[e.g.][]{krich,int_extr}.

In this paper we present new measurements for the scatter broadened quasar B\,2005+403, combining 
Very Long Baseline Interferometry  (VLBI) and flux density variability measurements obtained at various frequencies 
over the last decade. \object{B\,2005+403} is a flat spectrum quasar  
\citep[$S_{\rm 5\,GHz} = 2.5 - 3.5$\,Jy, $\alpha_{\rm 0.3/5\,GHz} = 0.3$, with $S \sim \nu^{\alpha}$;][]{spekt} at a redshift of
$z=1.736$ \citep{redshift}. It is located close to the Galactic plane at $l$\,=\,$76.82^\circ$, $b$\,=\,$4.29^\circ$ 
behind the \object{Cygnus super-bubble} region.
Earlier studies showed that interstellar scattering affects the VLBI image of the source, causing 
angular broadening at frequencies below 5\,GHz \citep{fey,mutel,desai}. 
The high value of the scattering measure ($ \text{SM}$\,=\,$\int_0^L C^{2}_{N}\left( s \right) ds $ \footnote{The 
Scattering Measure is the path integral over the coefficient $C^{2}_{N}$ of the electron 
density fluctuation wavenumber spectrum.}), 
derived for the line of sight of B\,2005+403 by \cite{fey},
reflects the strong influence of interstellar medium.

In order to further investigate the scatter broadening and the amount of scintillation in B\,2005+403,
we combined all available flux density monitoring data with our VLBI observations. 
The wide frequency coverage of our 
data facilitates a study of the interplay of the frequency dependent scattering effects, which dominates below 8\,GHz 
and the source intrinsic variability, which is best seen at and above 15\,GHz. At these higher frequencies
B\,2005+403 has not been intensively studied with VLBI before. The VLBI data cover a time range of 11 yrs 
(from 1992 -- 2003) and allow us to study and monitor the structural evolution of VLBI-core and pc-scale jet. The scatter
broadening observed in the VLBI images at the lower frequencies is measured and gives further constraints to the
properties of the intervening ISM. Additional parameters for the ISM are obtained from the IDV monitoring
of the source, performed with the Effelsberg 100 m telescope.

The paper is organized as follows. In Sect. 2, we present details of the VLBI observations and 
corresponding data reduction and introduce the long-term trends of the 
flux density variability. In Sect. 3 we discuss the source structure, the kinematics of the inner jet 
components and the possible location of the nucleus (VLBI core). In Sect. 3.4, we focus on the 
results of the low frequency VLBI observations and discuss the scatter broadening. In section 3.5 we discuss 
the long-term variability of B\,2005+403 and relate it to changes of the VLBI structure.
In Sect. 3.6 we discuss the implications from the observed intraday variability. Sect. 4 gives a concluding summary.

The following cosmological parameters are used throughout this paper:
\(H_{0}\)\,=\,\(71\ \text{km\ s$^{-1}$\ Mpc$^{-1}$,} \ \Omega_{\text{m}}\)\,=\,\(0.27
\ \text{and} \ \Omega_{\text{vac}}\)\,=\,\(0.73\).

\section{Data and data reduction}

\subsection{VLBI data}

We analyzed 17 VLBI datasets of B\,2005+403 from 14 different observing epochs obtained during 1992 -- 2003. 
These observations were performed with different VLBI networks (EVN, VLBA, Global) at various frequencies ranging 
from 1.6\,GHz to 43\,GHz. In our analysis we also included several data sets from the 2\,cm-survey
\citep[see][]{2cmsurvey} and one observation of the source in the MOJAVE survey \citep{mlister}. 
In Table  \ref{tab:data} we summarize the VLBI experiments.

After correlation at the VLBI correlators in Socorro (NRAO) or Bonn (MPIfR),
the VLBI data were calibrated and fringe-fitted using the standard procedures within AIPS 
(Astronomical Image Processing System) software package. 
The post-processing of the data invoked the usual steps of editing,
phase and amplitude self-calibration and imaging. These tasks were performed within AIPS and the
Caltech DIFMAP packages. For previously published datasets (from the 2\,cm-survey and
the MOJAVE survey and the datasets from epoch 2003.04), we reanalyzed the data. Starting
from pre-calibrated visibilities, we redid the self-calibration, imaging and Gaussian model fitting. 
For comparison of the results from this reanalysis with the original maps in the 2\,cm Survey we refer the reader
to \citet{2cmsurvey}.

In three VLBI experiments the source was also observed in full polarization. 
Column 4 of Table \ref{tab:data} summarizes the recorded polarization. In the polarization analysis, the 
feed leakage terms for the
antennas (D-terms) were determined in the usual way using the AIPS task ``LPCAL'' \citep{aips_pol}.
At 15\,GHz, we could use the University of Michigan (UMRAO) database 
to calibrate the orientation of the electric vector position angle (EVPA).
We verified that total and correlated flux density (in total intensity and in polarization)
matched within 10\,\% accuracy, excluding a large miscalibration of the EVPA owing to overseen
polarized large scale structure. At 22 and 43\,GHz, we had no possibility to calibrate the EVPA.
At these frequencies, the absolute orientation of the EVPA is therefore unknown.

\begin{figure*}[htbp!]
\centering
\includegraphics[width=15.5cm]{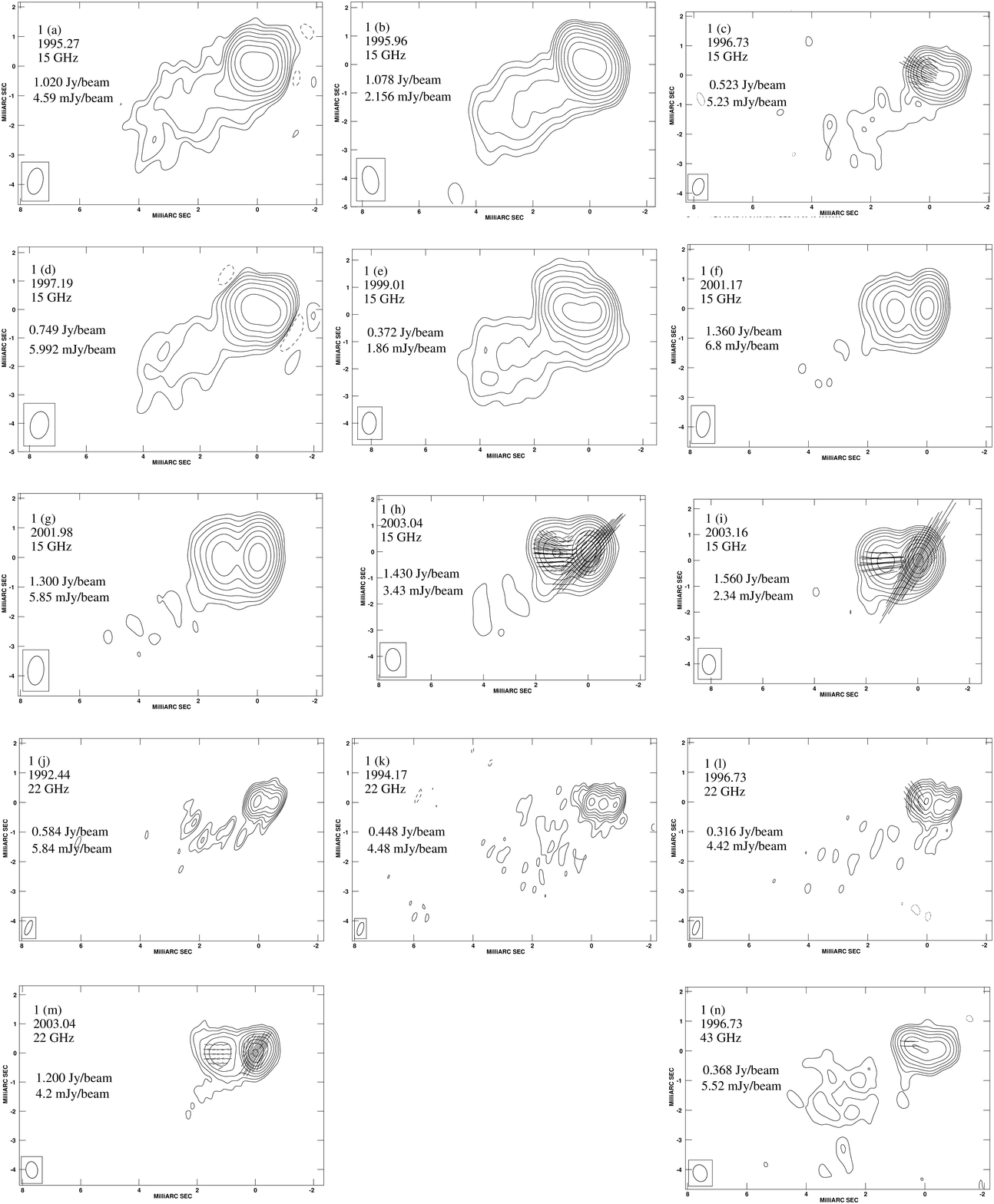}
\caption{VLBI CLEAN maps of B\,2005+403 at 15\,GHz, 22\,GHz, and 43\,GHz 
shown at the different observing epochs. Epoch, frequency, peak flux density and the lowest (positive) contour 
are given in the upper left corner of each image. Contours are in percent of the peak flux and increase by factors 
of two. The beamsize is shown in the bottom left corner of each image. 1 mas length of the superimposed 
polarization vectors corresponds to 12.5, 25, 10, 25, 31.3, 31.3 mJy/beam in images (h) to (n), respectively. 
(Images were created in AIPS.) }
\label{fig:im15}
\end{figure*}

In Fig. \ref{fig:im15} we present the results from the VLBI imaging, showing the
CLEAN maps obtained at 15\,GHz, 22\,GHz, and 43\,GHz. Figure \ref{fig:im15} also shows the polarization images 
with the electric vectors superimposed.

To facilitate measurement of structural changes, the VLBI maps were parameterized in the 
usual manner using Gaussian components. We fitted Gaussian components to the
calibrated visibility data using the DIFMAP program. The resulting modelfits are shown in a
separate table which is available online.
The table contains the following information: 
Col. 1 gives the observing epoch, frequency and VLBI array, Col.
2 identifies the VLBI components by letters and numbers, Col. 3 gives the flux density of each VLBI component ($S$), 
Col. 4 its radial distance $r$ relative to component C1, Col. 5 the position angle ($\theta$), 
Col. 6 the FWHM major axis ($a$), Col. 7
the ratio of minor to major axis ($b/a$), Col. 8 the position angle of the major axis ($\phi$), and Col. 9
the reduced $\chi^2$ to characterize the goodness of the modelfit. The positional information (Columns 4 and 5) 
are relative to the westernmost component, denoted as C1. It is further assumed that the position of C1 is stationary with time.
The measurements errors listed in the table were estimated from the internal scatter obtained during
the process of imaging and model-fitting and is described in more detail e.g. in \citet{cyg1}.
During the process of model fitting we reduced the number of independent variables by fitting only 
circular Gaussian components.
For those cases where circular Gaussian model fits gave not satisfactory results and produced unacceptable
large reduced $\chi^2$ values, elliptical Gaussian components were fitted. Most of the
22\,GHz, and 43\,GHz data were fitted best with elliptical Gaussians.

\begin{figure*}[htbp!]
\centering
\includegraphics[width=15.5cm]{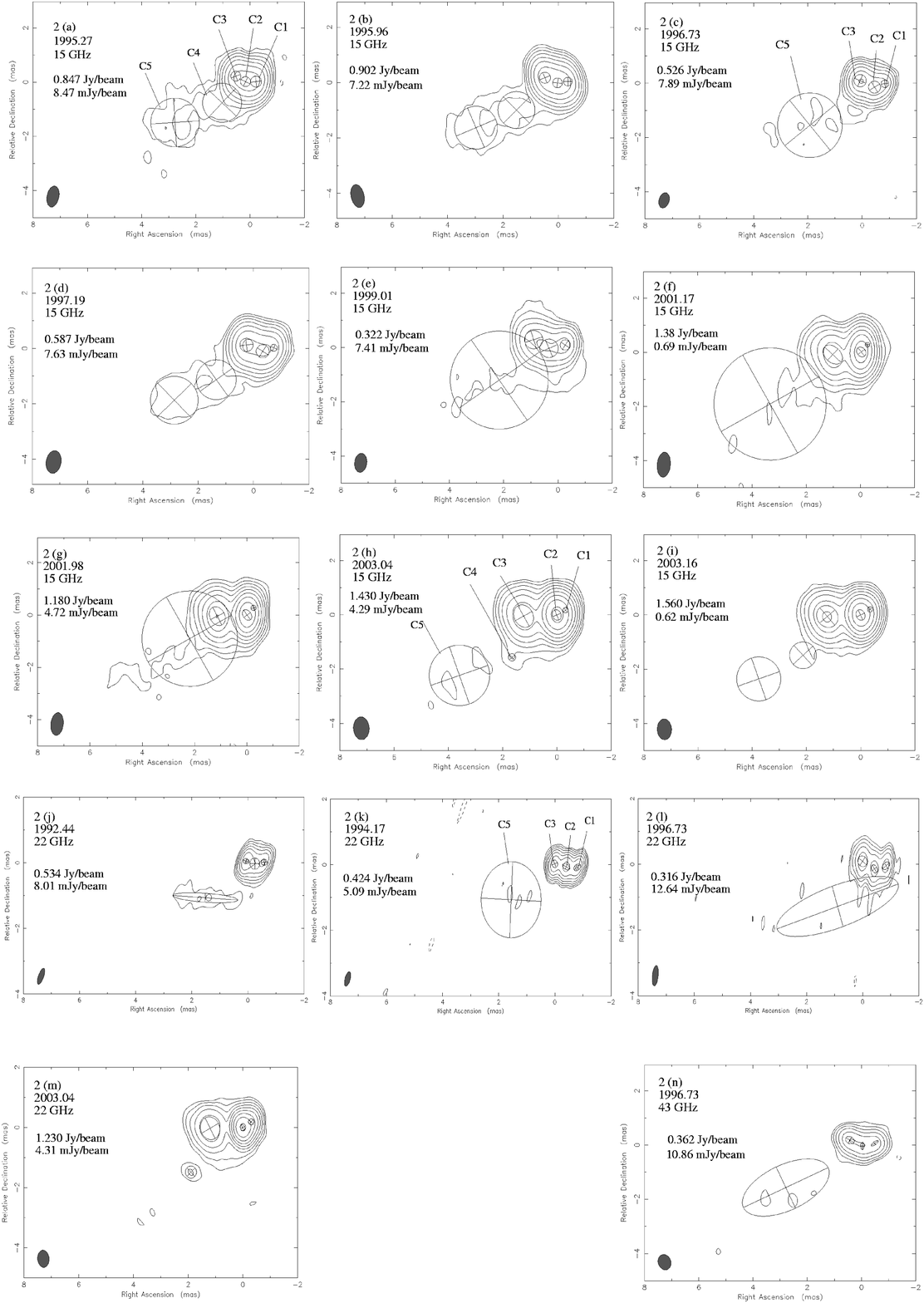}
\caption{Maps of Gaussian modelfits for B\,2005+403. Circles characterize positions and
sizes of the individual modelfit components. Contours result from the convolution with the observing beam. 
Epoch, frequency, peak flux density and the lowest (positive) contour level are given in the upper left 
corner of each image. Contours correspond to the modelfits, they  are in percent of the peak flux and 
increase by factors of two. As in Fig. \ref{fig:im15}, the VLBI observations are at 15\,GHz, 22\,GHz, 
and 43\,GHz. (Images were created in DIFMAP.)}
\label{fig:mod15}
\end{figure*}

In Figure \ref{fig:mod15} we show the VLBI images resulting from the Gaussian model fitting.
The circles seen in this figure denote for the positions and sizes of the individual model fit components.
The contour lines result from the convolution of the modelfit components with the observing beam and
the addition of the underlying (noise limited) residuals.

\begin{table*}
\centering
\caption{\label{tab:data}The epochs, frequencies and arrays of the
VLBI observations described in this paper. In Col. 4 we indicated the polarization  
mode of the observation. In Col. 5 references for previously published data are given.}
\begin{tabular}{ccccc}
\hline
Epoch & Frequency & Instrument & Pol. & Reference \\
\hline
\hline
1992.44 & 22\,GHz & EVN (8 stations) & LL & this paper \\
1994.17 & 22\,GHz & VLBA (8 stations) + VLA + EVN (4 stations) & LL & this paper \\
1995.27 & 15\,GHz & VLBA (2\,cm Survey) & RR & \cite{2cmsurvey} \\
1995.96 & 15\,GHz & VLBA (2\,cm Survey) & LL & \cite{2cmsurvey} \\
1996.73 & 43\,GHz & VLBA+EB & full & this paper \\
1996.73 & 22\,GHz & VLBA+EB & full & this paper \\
1996.73 & 15\,GHz & VLBA+EB & full & this paper \\
1996.82 & 5\,GHz & EVN (8 stations) & LL & this paper \\
1996.83 & 8\,GHz & EVN (8 stations) & RR & this paper \\
1997.19 & 15\,GHz & VLBA (2\,cm Survey) & LL & this paper \\
1998.14 & 1.6\,GHz & EVN (9 stations) & LL & this paper \\
1999.01 & 15\,GHz & VLBA (2\,cm Survey) & LL & \cite{2cmsurvey} \\
2001.17 & 15\,GHz & VLBA (2\,cm Survey) & LL & \cite{2cmsurvey} \\
2001.98 & 15\,GHz & VLBA (2\,cm Survey) & LL & \cite{kovalev} \\
2003.04 & 22\,GHz & VLBA & full & \cite{uwe_thesis} \\
2003.04 & 15\,GHz & VLBA & full & \cite{uwe_thesis} \\
2003.16 & 15\,GHz & VLBA (MOJAVE) & full & \citet{mlister,kovalev} \\ 
\hline
\end{tabular}
\end{table*}

\subsection{Total flux density monitoring}

B\,2005+403 is included in the flux density monitoring programs performed at
the University of Michigan Radio Astronomy Observatory (UMRAO) in the USA
and at the Mets\"ahovi Radio Observatory in Finland. This allows to
study the long-term flux density variability. In Figure \ref{fig:lc} we show the lightcurves
at 5\,GHz, 8\,GHz and 15\,GHz \citep{umrao} and at 22\,GHz and 37\,GHz \citep{metsehovi1, metsehovi2}.
The covered time range is 1990 -- 2004. A more detailed discussion of the variability and its
possible relation to the structural changes seen in the VLBI maps is given in Sect. 3.5.

In the context of ongoing systematic studies of Intraday Variable radio sources, B\,2005+403 was
observed with the Effelsberg 100\,m radio telescope of the Max-Planck-Institut f\"ur Radioastronomie.
During two observing runs especially dedicated to search for IDV in sources located
in the Cygnus region \citep[the observed sources are those studied by][]{fey,desai}, we
measured the flux density variability of B\,2005+403 at 6\,cm (4.85\,GHz) and 18\,cm (1.67\,GHz)
with the Effelsberg 100\,m radio telescope.
The observations were performed on November 30 and December 6, 2003. The flux densities were
measured adopting the method of repeated and subsequently averaged cross-scans. The measurement
method and the data analysis is described in more detail e.g. in \cite{reduc_idv}.
The duty cycle of the measurement resulted in $1-2$ flux density measurements
per source and hour. For calibration purposes, several steep spectrum sources were included and
observed with the same duty cycle as the target source
B\,2005+403. From these additional sources, the source \object{B\,2021+614} showed the least variability
and therefore could be used as secondary calibrator to monitor
and correct for residual short time-scale fluctuations of the antenna gain. The absolute
flux density scale was determined from repeated observations of the primary calibrators
\object{NGC\,7027}, \object{3C\,48} and \object{3C\,286} and the flux density scale
of \citep{abs1, abs2}.

\section{Results} 

\subsection{Kinematics}

In Fig. \ref{fig:im15} we show the VLBI CLEAN maps obtained at 15\,GHz, 22\,GHz, and 43\,GHz. 
The corresponding maps of the Gaussian modelfits are presented in Fig. \ref{fig:mod15}.

The maps of B\,2005+403 show an east-west oriented core-jet structure with 
several embedded components.  
The central 1\,mas region is best described by three Gaussian components C1, C2, and C3. 
The relative alignment of these 3 components seems to vary over the time spanned by the observations. In most
images, the relative alignment of C1 -- C3 suggests a slight bending to the north. Beyond 1\,mas core separation and
oriented more to the south-east, a faint and partially resolved region of diffuse and extended emission
is visible. It can be fitted by one or two Gaussian components of larger extend (C4 and C5).

\begin{figure}
  \resizebox{\hsize}{!}{
    \subfigure{\includegraphics[clip,width=\columnwidth]{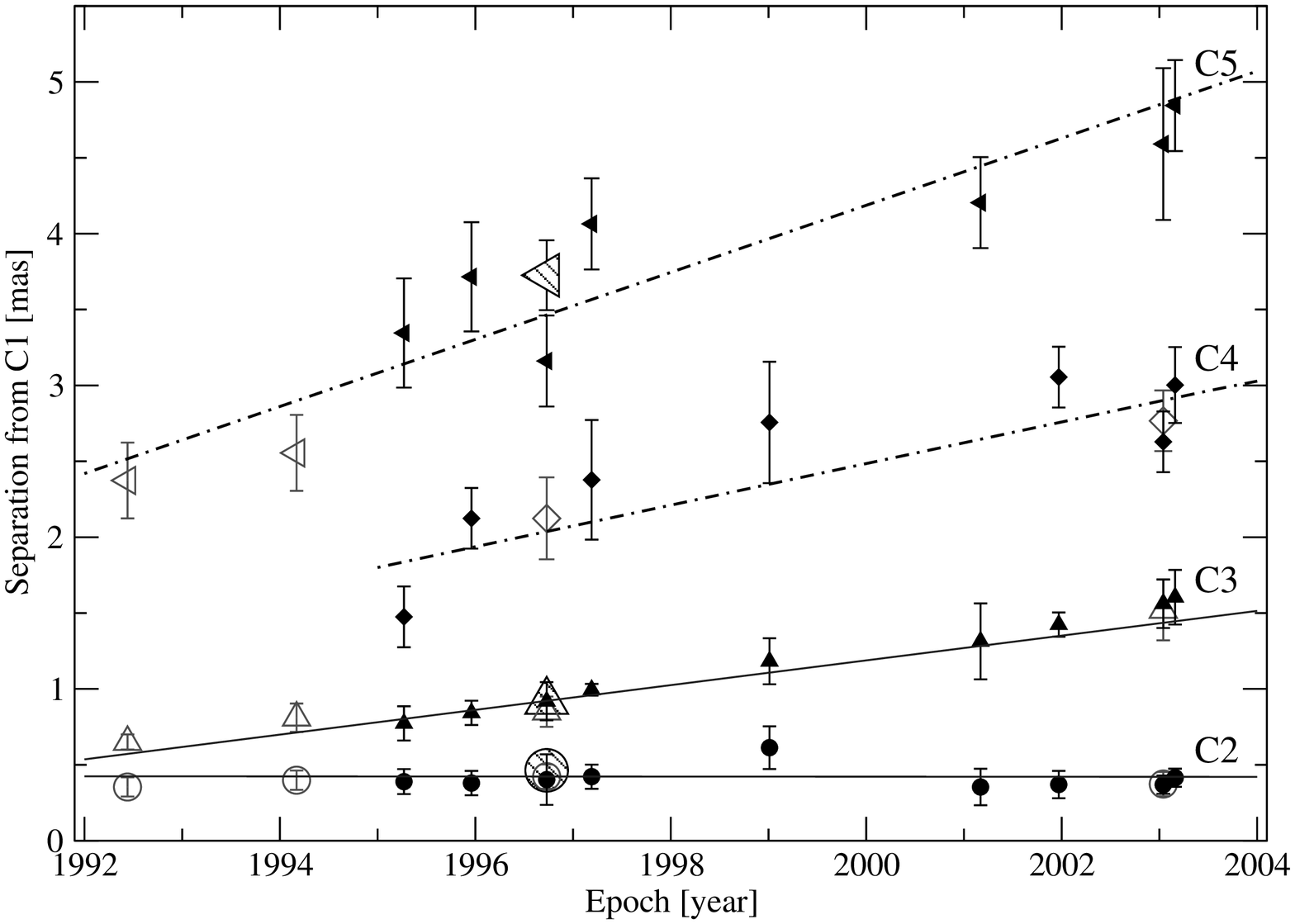}}}
  \resizebox{\hsize}{!}{
    \subfigure{\includegraphics[clip,width=\columnwidth]{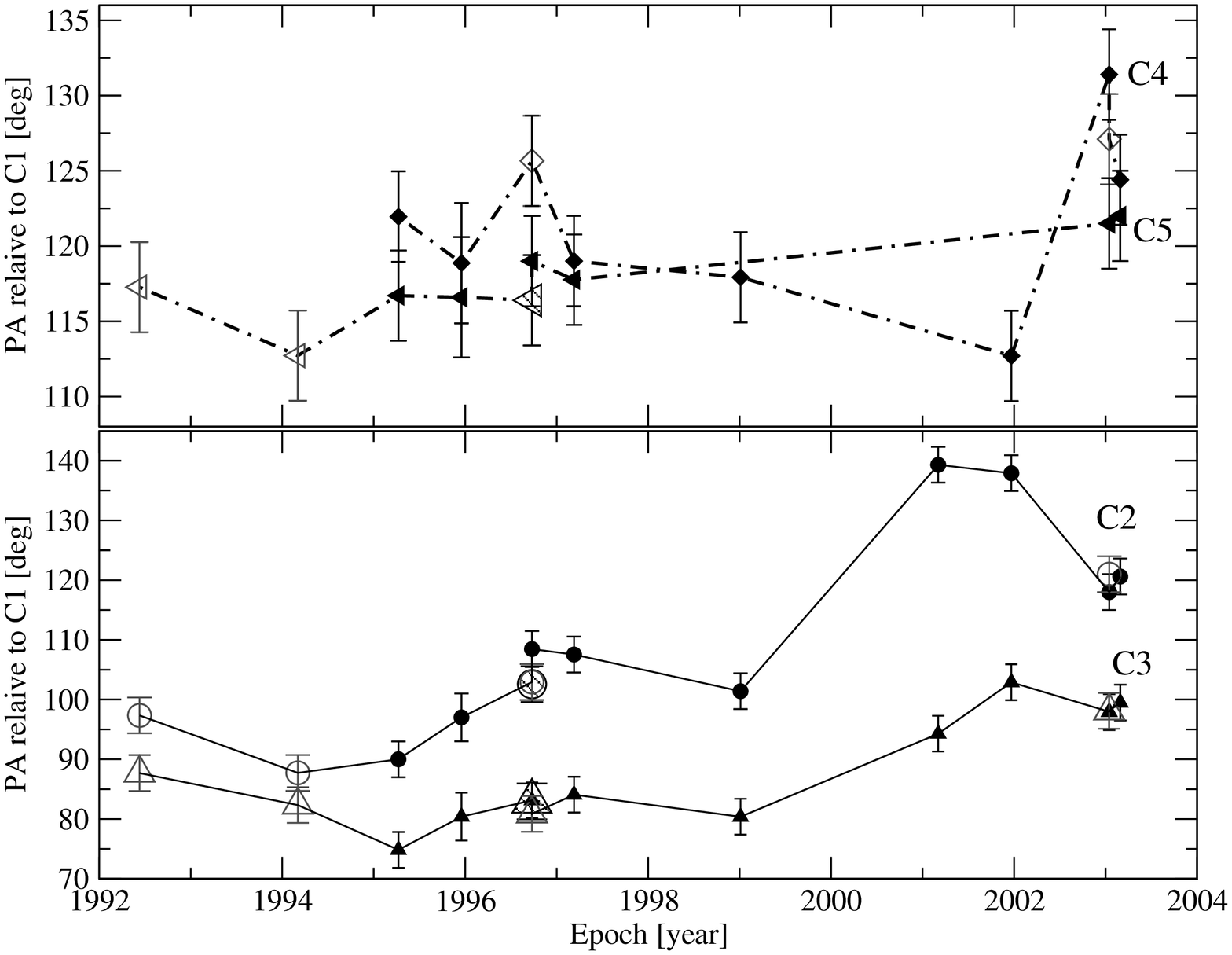}}}
  \caption{Top panel: core separation as a function of time for components: C2 (circles), C3 (triangles), 
    C4 (diamonds), C5 (left triangles).  Filling of a symbol 
    denotes the frequency of the observation: open for 22\,GHz, filled for 15\,GHz, larger and
    striped for 43\,GHz. Solid lines indicate linear regression to reliable identifications (C2 and C3).  
    The dashed-dotted lines show regression to tentatively suggested identifications. 
    Bottom panel: components position angle as a function of time. Same symbols and lines 
    are used as above. Lines are used just to guide the eye.}
  \label{fig:C1_year}
\end{figure}

In Fig. \ref{fig:C1_year} we plot the relative separation from C1 for the components C2 to C5 versus time.
Owing to its inverted spectrum, compactness and variability, we assume that C1 is the VLBI core 
and that its position is suitable for use as a stationary reference point (Sect. 3.2 for a discussion). 
In the figure, we also superimpose a straight line resulting from a linear fit to the motion of each component. 
The slope of each line measures the angular separation rate. These rates and the corresponding
apparent velocities are summarized in Table \ref{tab:kinematics}.

With the exception of component C2, which appears to be stationary within the measurement
errors, the components C3 to C5 move with apparent superluminal velocities. For these three components
we observe a systematic increase to their apparent velocities from 6.3\,c for component C3, to 9.9\,c
for C4 and 16.8\,c for C5. In the 2\,cm survey \citep{2cmsurvey}, B\,2005+403 is characterized by a 
two component model, with the secondary component moving at an apparent speed of $12.3 \pm 3.0$\,c.
Based on its core-separation, this component may be identified with a blending of components C4 and C5
in our analysis.

The minimum Lorentz factor $\gamma_{\text{min}}$ of the jet could be estimated from the maximum observed component
speed. Adopting the usual equations for

\[\gamma_{\text{min}}=\sqrt{1+{\beta_{\text{app}}}^2}\]  and the inclination angle of the jet, which leads
to maximum apparent speed 
\[\psi_{\gamma_{\text{min}}}=\cot^{-1}{\beta_{\text{app}}},\]
one obtains for component C5 $\gamma_{\text{min}} \geq 16.8 \pm 2.3$  and
$\psi_{\gamma_{\text{min}}} = 3.3^\circ \pm 0.5^\circ$.

\cite{doppl} calculated the Doppler boosting factor for a sample of active 
galactic nuclei using the total flux density monitoring data of the Mets\"ahovi observatory. 
For B\,2005+403, they report a variability Doppler boosting factor of $\delta_{\text{var}}$\,=\,$8.63$,
which translates to $\delta_{\text{var}}$\,=\,$10.9$, for the cosmological parameters used in this paper.
We note that because of the light travel time argument used by \cite{doppl}, the derived Doppler-factor
places a lower limit to the true Doppler-factor.
With the proper motions obtained from VLBI, we may calculate the viewing angle and the Lorentz factor of the jet:
\[\psi_{\text{var}}=\arctan{\frac{2\beta_{\text{app}}}{\beta_{\text{app}}^2+\delta_{\text{var}}^2-1}}\]
\[\gamma_{\text{var}}=\frac{\beta_{\text{app}}^2+\delta_{\text{var}}^2+1}{2\delta_{\text{var}}}\] 
$\psi_{\text{var}} \leq 4.8^\circ \pm 0.9^\circ$ and $\gamma_{\text{var}} \geq 18.4 \pm 3.5$.

The apparent acceleration observed for components C2 to C5 may either result from an intrinsic acceleration
of the jet, or from a systematic bending of the jet, or from a combination of both effects.
As $\psi_{\gamma_{\text{min}}}$ and $\psi_{\text{var}}$ both are very small, a jet bending of less than 
a few degrees could easily explain the observed increase of the apparent velocities. To investigate
this further, we plot in the lower panel of Fig. \ref{fig:C1_year} 
the apparent variations of the position angle of the radius vector for each component.
The plot indicates that component C2 and C3 move along similar and spatially bent trajectories
with position angle changes in the range of $\sim 20-40^\circ$. Between 1992 and 2003 the position 
angle difference between  C2 and C3 increases, indicating still similar but spatially offset paths. 
For components C4 and C5 the situation
is less clear, as much smaller variations of the position angles are observed and the scatter in the data 
is larger. We therefore conclude that at least the components C2 and C3 move on non-ballistical and spatially
bent trajectories.

\begin{figure}
  \resizebox{\hsize}{!}{
    \includegraphics[clip,width=\columnwidth]{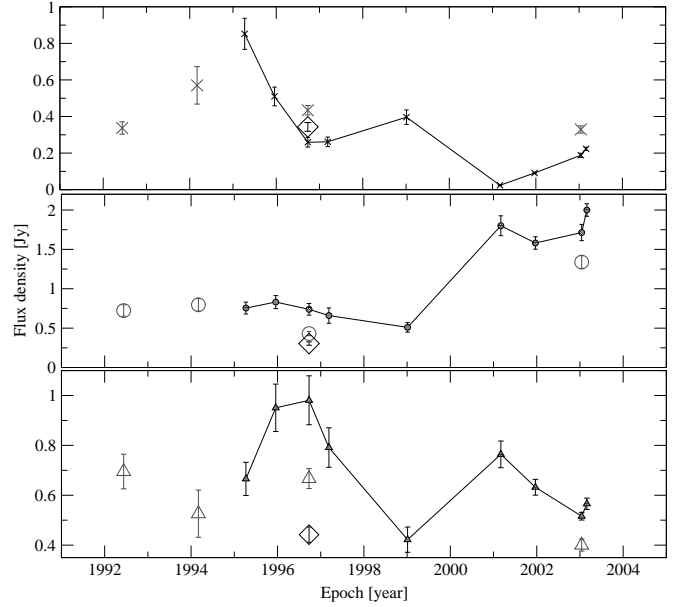}} 
  \caption{Flux density evolution of individual jet components C1, C2 and C3 is shown. C1 is assumed to be the core. 
    Open grey symbols denote observations at 22\,GHz, filled black symbols at 15\,GHz and open diamonds represent the observation at 43\,GHz.}
  \label{fig:flux_year}
\end{figure}

The variation of the flux density of the individual jet components is plotted versus time in Fig. \ref{fig:flux_year}. 
Since most data are available at 15\,GHz, we connect the flux density measurements at this frequency
with lines. The variations of the flux density of components C1, C2 and C3 appear completely uncorrelated,
with the flux density of C1 mainly decreasing, that of C2 increasing and the flux density of C3 showing
some oscillatory behavior with at least two local maxima. In Section 3.5 we will discuss the combined
flux density variability of the VLBI components with the variations seen in the total flux density of the source.

\begin{table}
\caption{\label{tab:kinematics}The apparent speed and the corresponding apparent proper motion of components C2, C3, C4 and C5.}
\begin{tabular}{ccc}
\hline
Component & Apparent speed & Apparent proper motion \\
 &  $\text{\,mas yr}^{-1}$ & in units of $c$ \\
\hline
\hline
C2 & $\left( -0.3 \pm 4.0 \right) 10^{-3}$ & $\le 0.31 $\\
C3 & $0.082 \pm 0.006$ & $6.25\pm 0.46$ \\
C4 & $0.14 \pm 0.03$ & $9.91 \pm 2.12$ \\
C5 & $0.22 \pm 0.03 $ & $16.77 \pm 2.29$ \\
\hline
\end{tabular}
\end{table}

\subsection{The position of the VLBI core}

From our simultaneous multi-frequency VLBI observations the identification of the VLBI core and the measurement
of individual component spectra is facilitated.
For those components, where quasi-simultaneous flux density measurements were available at different frequencies, we calculated 
the spectral indices and show in Table \ref{tab:spekt} the results.
In 1996.73 and 2003.04, component C1 shows between 15\,GHz and 22\,GHz an inverted spectrum. Between  22 and 43\,GHz, the
spectrum steepen. Components C2 and C3, both show an optically-thin synchrotron spectrum, with similar spectral
indices of $\sim -0.7$ consistently measured in both observations. The spectral shape of C1 can be explained
via synchrotron-selfabsorption. This, its flux density variability and the high compactness seen in the modelfits
indicate, that C1 has to be identified with the synchrotron-selfabsorbed jet base, which commonly is called the VLBI core.

We note that despite its optically thin spectrum, component C2 shows the most pronounced flux density variability
(see  Figure \ref{fig:flux_year}). This is somewhat atypical, as pronounced flux density variability is
usually seen in jet components with much flatter spectra. The apparent flux density increase
of $\Delta S_{\rm 15 GHz} = 1.3$\,Jy in $\Delta t_{\rm var} = 2.16$\,yrs leads via the light travel time argument 
to an apparent brightness temperature of $T_\text{B} =2.2 \cdot 10^{12}$\,K, which apparently violates the inverse Compton limit. 
The calculation lead to a lower limit for the Doppler-factor of $\delta \geq 3.6$, which would be required to
reduce the observed brightness temperature to $10^{12}$\,K.
It is therefore very likely that the variability is mainly caused by differential Doppler boosting 
in combination with motion along a curved path.
The relatively large variability and the apparent stationarity of C2 strongly indicate motion more or less directly towards the
observer, i.e. an angle to the line of sight close to zero. A similar interpretation to explain apparent stationarity
has been made also for other sources, i.e. for \object{4C 39.25} \citep{4c39}.

On the basis of their spectral shapes with a steep spectral index of $-0.7$, we rule out that components 
C2 or C3 can be identified with the VLBI core. An implication of this is, that B\,2005+403 is a one-sided
core-jet source and that it exhibits no counter-jet. The lack of a counter-jet is usually explained as a
consequence of Doppler-boosting, consistent with our observations.

The spectral shape of C1 indicates the existence of a spectral maximum near 22\,GHz (Table \ref{tab:spekt}).
If this is identified with the synchrotron-turnover of a homogeneous synchrotron self-absorbed component,
we can estimate the magnetic field strength using the measured size of component C1: $0.34$\,mas. 

Following \citep[e.g.][]{magnet}, the magnetic field strength is:
\[B=10^{-9} \delta \frac{b(\alpha) \theta^{4} \nu^5}{S^{2}(1+z)} \text{\,T}\]
where $\delta$ is the Doppler factor, $\theta$ is the component size in units of mas, $\nu$ is the 
turnover frequency in GHz, $S$ (in Jy) is the flux density at the turnover frequency and  $b(\alpha)$
is a spectral index dependent tabulated parameter \citep[see Table 1 in][]{magnet}. Here we adopt $\alpha=-0.35$. 
This yields for C1 $B$\,=\,$1.7\delta\cdot10^{-4} \text{\,T}$.

\begin{table}
\centering
\caption{\label{tab:spekt}Spectral indices of the components C1, C2, and C3 ($S \sim \nu^{+\alpha} $)}
\begin{tabular}{c|*3{>{$}c<{$}}}
Epoch & \text{C1} & \text{C2} & \text{C3} \\ 
\hline \hline 
1996.73 & \alpha_{15}^{22}=1.34 & \alpha_{15}^{43}=-0.70 & \alpha_{15}^{43}=-0.70 \\
& \alpha_{22}^{43}=-0.35 & & \\
\hline
2003.04 & \alpha_{15}^{22}=1.46 & \alpha_{15}^{22}=-0.65 & \alpha_{15}^{22}=-0.67 \\
\hline
\end{tabular}
\end{table}

\subsection{Polarization}

In Fig. \ref{fig:im15} the polarization vectors are superimposed on the VLBI maps for those epochs 
where we have polarization data. The lengths of the vectors represent the 
polarized flux density. The polarized flux and the polarization angles derived from the VLBI observations are 
displayed in Table \ref{tab:pol}. Owing to the lack of absolute polarization angle calibration at 22\,GHz, and 43\,GHz,
the position angles of the electric vectors are arbitrary at these frequencies. 
The measured polarization characteristics at 15\,GHz are in agreement 
with the UMRAO single dish measurements.

Component C2 was unpolarized in 1996.73, however this was the most prominent polarized feature in 2003.04 and 2003.16, 
with comparable polarized flux density at 15\,GHz and 22\,GHz. The polarized flux density of C3 decreased through all the epochs
and it also decreased by increasing frequency.

The polarization angles of C2 did not change significantly with time. The polarization angle of C3 
changed by $\approx 15^{\circ} \text{ and } 40^{\circ}$, at 15\,GHz, and  22\,GHz respectively, 
from 1996.73 to 2003.04. In 2003.04 and 2003.16 it remained at the same value, roughly parallel 
with the jet direction implying that the magnetic field is 
perpendicular to the jet, as one expects from an optically thin component.  

The absolute angle of polarization is unknown at 22\,GHz and 43\,GHz however the angles in all epochs for both components are
not significantly different from the calibrated EVPA at 15\,GHz (one exception is 1996.73 at 22\,GHz). This is consistent 
with the low Rotation Measure\footnote{RM is an integration
of magnetic field strength multiplied by the electron density along the line of sight from the 
source to the observer: $RM=812\int_0^L n_\text{e} B_{||} dl \text{\,rad\,m}^{-2}$, where $B_{||}$ is the 
parallel component of the magnetic field in mG.} (RM) values determined by \cite{zavala}.

\begin{table}
\centering
\caption{\label{tab:pol}Polarization characteristics of B\,2005+403 obtained from the VLBI maps. Col. 1 gives the epoch, Col. 2 the frequency of the observation.  Col 3 gives the name of the component. Col. 4 gives the polarized flux density in mJy, Col. 5 the polarization angle in degrees. The EVPA is not calibrated the 22\,GHz and 43\,GHz datasets; the given values are arbitrary.}
\begin{tabular}{|c|*1{>{$}c<{$}}|c*3{|>{$}c<{$}}}
\hline
Epoch & \nu \text{ [GHz]} & ID & \text{P [mJy]} & \chi (^\circ) \\ 
\hline \hline 
2003.04 & 15 & C2 & 66 \pm 14 & 131\pm3\\
2003.04 & 22 & C2 & 74 \pm 14 & 131\pm3\\
2003.16 & 15 & C2 & 67 \pm 14 & 121\pm6\\
\hline
1996.73 & 15 & C3 & 42 \pm 5 & 75\pm8 \\ 
1996.73 & 22 & C3 & 35 \pm 4 & 49\pm8\\
1996.73 & 43 & C3 & 13 \pm 3 & 90\pm4\\ 
2003.04 & 15 & C3 & 38 \pm 4 & 90\pm4\\ 
2003.04 & 22 & C3 & 21 \pm 3 & 90\pm4\\
2003.16 & 15 & C3 & 28 \pm 5 & 90\pm5\\
\hline
\end{tabular}
\end{table}

\subsection{Low frequency VLBI observations} \label{scat}

The central part (denoted as M1 in Table \ref{tab:model15}) of B\,2005+403 cannot be resolved at
1.6\,GHz, 5\,GHz, and 8\,GHz. However, at 5\,GHz and 8\,GHz two and one extended features appear in the jet 
direction, towards south-east (M2 and M3). M2 can be tentatively identified with C5 in the 15\,GHz and 22\,GHz data. 

\begin{figure}
  \rotatebox{-90}{
	\includegraphics[clip,width=\columnwidth]{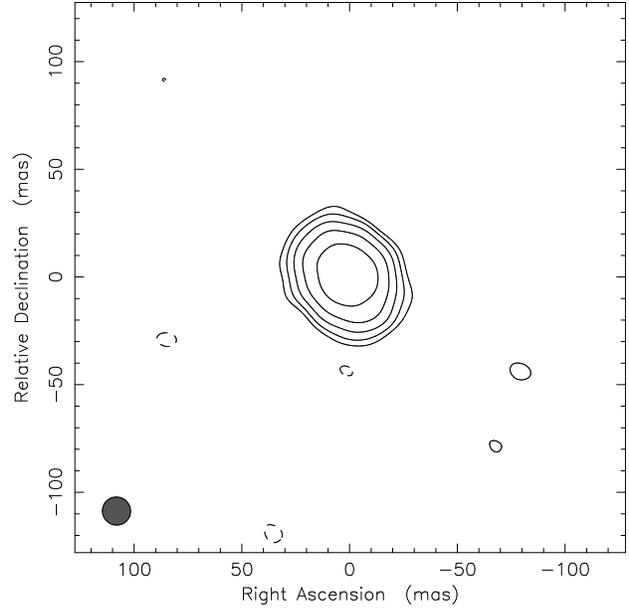}}
  \caption{Clean map obtained in EVN observation (at 1.64 GHz) in 1998.14. The peak flux density is 0.464 Jy/beam. 
    The contour levels are 4.64 mJy/beam $\times$(-3.2, 3.2, 6.4, 12.8, 25.6, 51.2). The beam FWHM is $13$ mas, 
    in contrast to the apparent angular size of the source of $36$ mas.}
  \label{fig:bro}
\end{figure}

Our low frequency VLBI observations at 1.6\,GHz, 5\,GHz and 8\,GHz confirm the effect of angular 
broadening reported earlier by  \cite{mutel}, \cite{fey} and \cite{desai}. In Figure \ref{fig:bro}
we show 1.6\,GHz map obtained with the EVN in 1998.14. The image illustrates the effect of scatter
broadening: the measured source size (36\,mas) is three times larger than the size of the observing beam ($13$\,mas). 
The possible elongation seen at this map in the position angle of $\sim 37^{\circ}$ may indicate anisotropy 
in the scattering.

\begin{figure}
	\includegraphics[clip,width=\columnwidth]{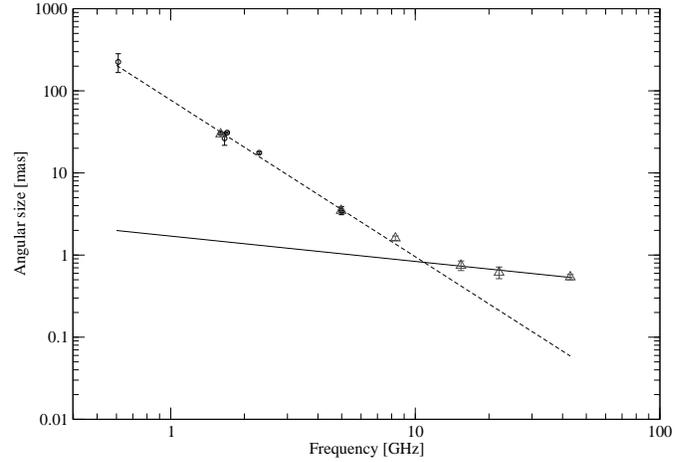}
  \caption{The measured angular size plotted versus observing frequency. The dashed line represents 
    a power law fit to the data in the range of 0.6\,GHz to 8\,GHz. The slope of the line is $-1.91\pm 0.05$. 
  The solid line represents a power law with a slope of $-0.31$, fitted to the data points at the high frequencies 
  in the range of 15\,GHz to 43\,GHz. Circles denote data from the literature, triangles data from this paper.}
  \label{fig:logplot}
\end{figure}

Similar to \cite{fey}, we analyze the measured total angular source size as a function of the observing frequency.
For this, we fitted the visibility data of B\,2005+403 at all available frequencies with one Gaussian component.
In Figure \ref{fig:logplot} we plot the FWHM size of this component versus frequency, using our data obtained at
1.6\,GHz, 5\,GHz, 8\,GHz, 15\,GHz, 22\,GHz, and 43\,GHz and including the previous measurements from the
literature \citep{fey,mutel,desai}, the corresponding values are tabulated in Table \ref{tab:scat}.
The figure clearly shows a deviation from a simple power-law. In order to quantify the scatter broadening, which
dominates at lower frequencies, we fitted a power law to the data in the frequency range
of $0.67 - 8$\,GHz. In this range we obtain the following result:

\(\theta\)\,=\,\((77.1 \pm 4.0)\cdot (\nu/1\text{\,GHz})^{-(1.91 \pm 0.05)}\)

The dashed line in Figure \ref{fig:logplot} shows this fit. We note that the slope
does not change significantly, if the data points at 8\,GHz are excluded from the fit. In this
case one obtains: $\theta =(77.7 \pm 4.0) \cdot (\nu/1\text{\,GHz})^{-1.92 \pm 0.06}$. In both cases the slope of the
power-law differs significantly from the slope of $-2.2$ one would expect for Kolmogorov type 
density fluctuations in the ISM. 

Above 8\,GHz the source is significantly larger than the extrapolated scattering size. 
This indicates that towards higher frequencies scattering effects are less dominant and that
the intrinsic structure of the sources shines through. The differences between the 
extrapolated scattering size and the measured source size are as follows:
$\Delta\theta_\text{15\,GHz}=(0.4\pm0.1)$\,mas, $\Delta\theta_\text{22\,GHz}=(0.4\pm0.1)$\,mas,
and $\Delta\theta_\text{43\,GHz}=(0.4\pm0.1)$\,mas.
For the three size measurements above 8\,GHz, we fit another power law, which to first order can be used 
to characterize the frequency dependence of the intrinsic source size:
$\theta_{\rm int}$\,=\,$(1.7\pm 0.4)\cdot (\nu/1\text{\,GHz})^{-(0.31 \pm 0.07)}$ mas.
The solid line in Fig. \ref{fig:logplot} shows this fit. With this curve we are able to
obtain an upper limit of the intrinsic source size also at lower frequencies, were direct size
measurements with VLBI are not possible.

Because of its heavily scattered line of sight, B\,2005+403 was incorporated by Cordes and Lazio in their model of 
describing the distribution of the Galactic free electrons \citep{gal_model}. To account for the regions of 
intense scattering, they included ``clumps'', which are regions of enhanced electron density and/or 
electron density fluctuation. Adopting the scattering measure 
of \cite{fey} 
thickness $d\approx 18 \text{\,pc}$ located at a distance of $L\approx2.35 \text{\,kpc}$ was derived. In their model,
a Kolmogorov spectrum of the interstellar turbulence was assumed.

At a given frequency, the size of the scattering disk can be estimated via the deconvolution formula:
$\theta_{\rm scat} = \sqrt{\theta_{\rm obs}^2 - \theta_{\rm int}^2}$. For the scattering size at 1\,GHz,
we obtain $\theta_{\text{1\,GHz}}$\,=\,$(77.1 \pm 4.0)$\,mas. Following \cite{tc93} and assuming Kolmogorov turbulence,
we obtain for the scattering measure:

\[\text{SM}=\left(\frac{\theta_\text{scat}}{128\text{\,mas}}\right)^{5/3}\cdot\nu_\text{GHz}^{11/3}.\]

The derived value of 
$\text{SM}$\,=\,$(0.43 \pm 0.04) \text{m}^{-20/3} \text{\,kpc}$ is in good agreement and consistent with the
previous measurement of \cite{fey}, however now has a smaller uncertainty. Our slightly lower scattering
measure implies either a somewhat reduced electron density fluctuation than that given in the 
\cite{gal_model} model or a shorter path length through
the screen. For the electron density fluctuation as given in \cite{gal_model}, 
we formally obtain a path length of 11.7\,pc, instead of the 18\,pc mentioned above.

\cite{em} reports on the measurements of H$\alpha$ intensity towards the Cygnus region. They 
studied lines of sight towards extragalactic radio sources which are affected by interstellar 
scattering. In the direction of B\,2005+403, the reported H$\alpha$ intensity 
is 130 Rayleigh, which corresponds to an emission measure of $EM=293.8\text{\,cm}^{-6}\text{\,pc}$ 
(assuming an electron temperature of $8000$\,K). The EM corresponds to the path integral of the squared electron density
along the line of sight. For a thin screen it can be 
approximated by: $EM=n_\text{e}^2 \cdot d$.  Assuming that the probed HII region causes the 
scatter broadening using equation (16) and (18)  from \cite{gal_model} we calculated 
from the EM and SM values the outer scale to be $l_0=25.5$\,AU, and the fractional variance of
the electron density to be $6.9$. Using the thickness of 11.7\,pc 
the electron density can be calculated: $n_\text{e}=5.0 \text{\,cm}^{-3}$.

The exponent of the power-law of the angular broadening ($-1.91\pm0.05$) indicates a wavenumber
spectrum of the electron density turbulence, which is flatter than the usually assumed Kolmogorov type
turbulence with spectral slope of $-2.2$. This can be produced in a medium consisting of random superpositions
of discontinuities such as a collection of shock fronts \citep{rick_uj}. 
\cite{scint} tabulate numerical estimates of scintillation parameters for different power-law spectra of
electron density fluctuations. We used the equation with a power law index of -2 from Table 1 of \cite{scint} 
that describes most adequately our angular broadening observation. From this we obtain 
$C_N^2 d$\,=\,$15.16 \text{\,m}^{-20/3} \text{\,kpc}$. With a thickness of the screen in the range of $d=12-18$\,pc
we obtain a range for $C_N^2$\,=\,$(842...1266) \text{\,m}^{-20/3}$. However, we note that this is not 
an entirely consistent approach, as \cite{gal_model} used a Kolmogorov spectrum.

From the power-law fitted to the source size at high frequencies, an upper limit for 
the intrinsic source size and hence lower limit for the brightness temperature can be calculated.
At 1.6\,GHz, 5\,GHz, and 8\,GHz the intrinsic source sizes are $\theta_\text{int}\le(1.5\pm0.6)$\,mas, 
$\theta_\text{int}\le(1.0 \pm0.3)$\,mas, and $\theta_\text{int}\le(0.9\pm0.2)$\,mas, respectively. 
The corresponding lower limits to the brightness temperatures are $T_\text{B}\ge(0.6\pm0.5)\cdot10^{12}$\,K, 
$T_\text{B}\ge(0.14\pm0.09)\cdot10^{12}$\,K, and $T_\text{B}\ge(0.08\pm0.04)\cdot10^{12}$\,K. These numbers
are in accordance with typical brightness temperatures measured in VLBI \citep[e.g.][]{kovalev}
and neither strongly violate the inverse Compton limit nor indicate excessive Doppler-boosting.

\begin{table}
\centering
\caption{\label{tab:scat}The measured angular sizes of B\,2005+403 at different frequencies. Col. 1 shows the observing frequency, Col.2 the observing epoch, Col. 3 the major axis size of the fitted Gaussian component. In Col. 4 references for previously published data are given.}
\begin{tabular}{|*1{>{$}c<{$}}|c*1{|>{$}c<{$}}|c|}
\hline
\nu \text{ [GHz]} & Epoch & \theta \text{ [mas]} & Reference \\ 
\hline \hline 
0.61 & Oct 1986 & 225 \pm 58 & \cite{fey} \\
1.6 & 1998.14 & 29.8 \pm 0.5 & this paper \\
1.66 & Mar 1986 & 26.4 \pm 4.7 & \cite{fey} \\
1.7 & Jan 1997 & 31 \pm 0.8 & \cite{desai} \\
2.3 & Jan 1997 & 17.6 \pm 0.5 & \cite{desai} \\
4.99 & Oct 1985 & 3.5 \pm 0.4 & \cite{fey} \\
5 & 1996.82 & 3.5 \pm 0.1 & this paper \\
5 & Jan 1997 & 3.4 \pm 0.1 & \cite{desai} \\
8 & 1996.83 & 1.6 \pm 0.1 & this paper \\
15 & 1996.73 & 0.8 \pm 0.1 & this paper \\
22 & 1996.73 & 0.6 \pm 0.1 & this paper \\
43 & 1996.73 & 0.5 \pm 0.04 & this paper \\
\hline
\end{tabular}
\end{table}

\subsection{Long-term flux density variations} \label{fluxvar}

\begin{figure*}
\centering
\includegraphics[width=17cm]{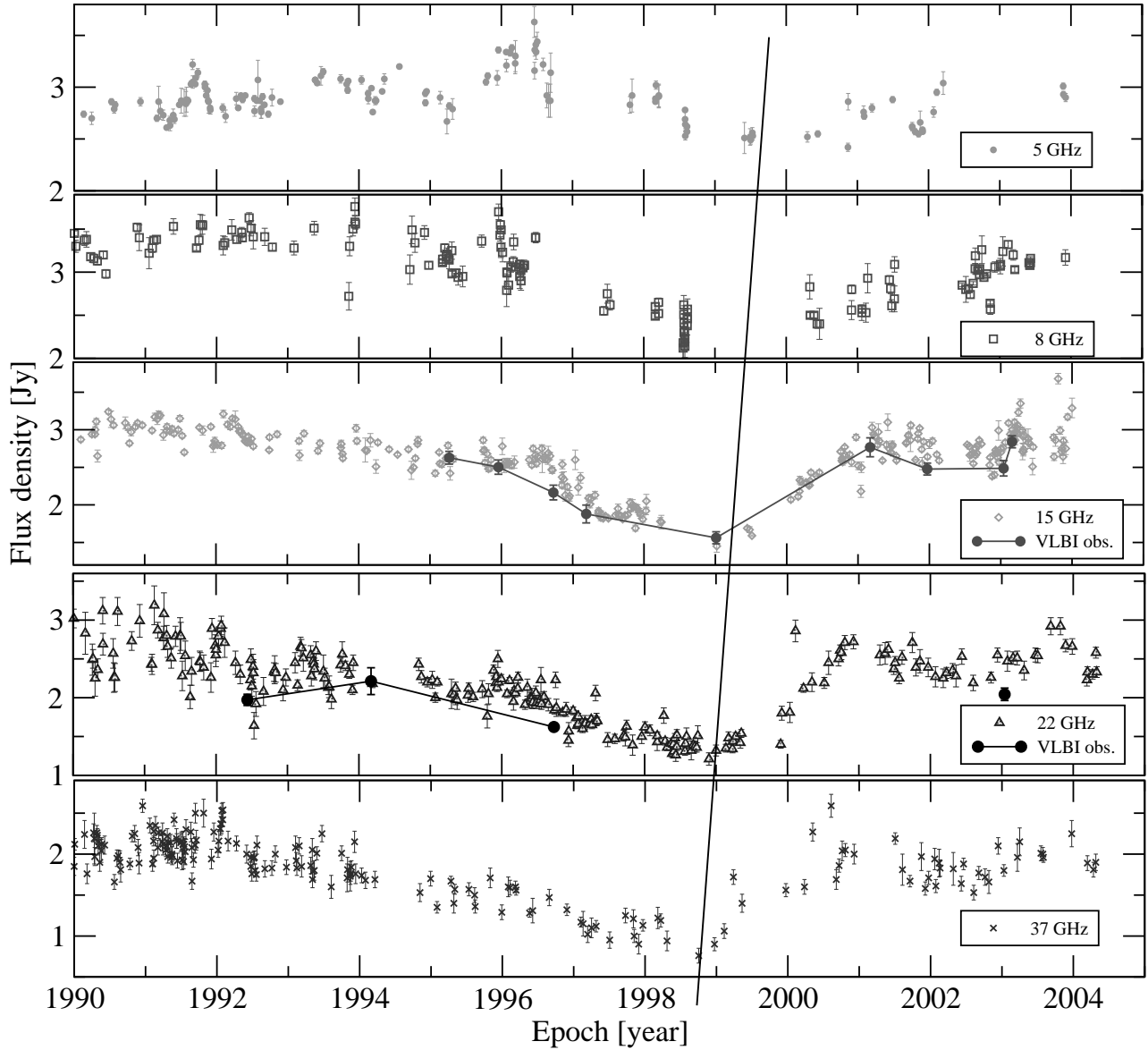}
\caption{Light-curves of B\,2005+403 at 5\,GHz, 8\,GHz, 15\,GHz, 22\,GHz, and 37\,GHz from the
monitoring data of the University of Michigan Radio Astronomy Observatory, and
the Mets\"ahovi 14 meter radio telescope. The straight line connects the minima. The dark circles
and lines denote the sum of flux density of the modelfit components from the VLBI observations at 15\,GHz and 22\,GHz.}
\label{fig:lc}
\end{figure*}

In Figure \ref{fig:lc} we show the radio light-curves of B\,2005+403 at 5\,GHz, 8\,GHz, 15\,GHz,
22\,GHz, and 37\,GHz during the time interval covered by our VLBI observations. The data between 5\,GHz
and 15\,GHz are from the Michigan (UMRAO) monitoring program, the data at 22\,GHz and 37\,GHz are
from the Mets\"ahovi monitoring.

The data clearly show the longterm variability of the source. A feature of
particular interest is the apparent decrease of the flux density observed between 1992
and 1999. Between 1999 and 2001, the flux increased again. This flux density 'trough' is best seen
and more pronounced at the higher frequencies.  At 5\,GHz, the decline
begins at 1996.5 and the flux density drops by 16\,\% reaching its lowest value around 1999.4.
At 8\,GHz, a flux density decrease of 30\,\% is observed during approximately the
same time interval. At 15\,GHz, the flux density drops by 47\,\% from 1996.6
till the beginning of 1999. At the two highest observed frequencies (22\,GHz and
37\,GHz), the decrease is even more pronounced: 54\,\% and 60\,\% respectively.
This systematic frequency dependence of the variability amplitude is accompanied by 
a systematic and frequency dependent shift of the time of the flux density minimum.
It is obvious that the flux density minimum and the subsequent rise of the flux density
appear earlier at the higher frequencies. In  Fig. \ref{fig:lc} we illustrate this frequency dependence 
by a solid line, which is not a fit and only meant to guide the eye.

To quantify the variations at the different frequencies we use the modulation index, defined as follows:
\[m=100 \frac{\sigma}{\langle S\rangle}\]\,\%,
were $\sigma$ denotes for the rms variation and $\langle S\rangle$ for the mean flux density. The mean flux
density and the corresponding variability indices are summarized in Table \ref{tab:m_index}. It is seen
that at 22\,GHz the modulation index is largest.

\begin{table}
\centering
\caption{\label{tab:m_index}The mean flux density, the standard deviation and the modulation indices 
for the flux density variability of B\,2005+430 at different frequencies between 1990 and 2004.}
\begin{tabular}{|*4{>{$}c<{$}|}}
\hline
\text{Frequency (GHz)} & \langle S\rangle \text{\,(Jy)} &  \sigma  & \text{m (\%)} \\ 
\hline \hline 
5 & 2.86 \pm 0.02 & 0.217 & 7.61 \\
8 & 3.03 \pm 0.02 & 0.369 & 12.15 \\
15 & 2.63 \pm 0.02 & 0.374 & 14.24 \\
22 & 1.95 \pm 0.03 & 0.500 & 25.67 \\
37 & 1.84 \pm 0.03 & 0.363 & 19.81 \\
\hline
\end{tabular}
\end{table}

In a number of AGN, short time flux density troughs are explained via 
an occultation by ISM clouds moving through the line of sight. This are the so called
``extreme scattering events'' \citep[ESE, see][]{ese}. Depending on size, distance and
relative velocity of the scatterer, timescale between a few days, months and perhaps even
years are possible \citep{cimo,pohl}.
Since B\,2005+403 is located in a region
of the sky with prominent scattering effects, the interpretation of the observed flux density trough 
via such an extreme scattering event provides an attractive possibility for an interpretation. However, 
the frequency dependence and variability time scales seen in B\,2005+403 are different than those 
expected for an ESE. Typical ESE events show more pronounced and more rapid
variations towards longer wavelengths, opposite to what is observed in B\,2005+403. Also the frequency
dependence of the flux density minimum observed in the trough of B\,2005+403 is not consistent
with an ESE event, which owing to the geometrical occultation should not show any frequency shifts.

The observing dates of the low frequency VLBI observations performed in 1996.8 at 5\,GHz and 
1998.1 at 1.6\,GHz coincide with the times of the decrease of the total flux density. The observed
angular source sizes at these times, however, are not significantly different than the sizes
observed previously (1985 \& 1986) by \cite{fey} and long before the flux density trough.
This indicates that the amount of the observed angular broadening is persistent and therefore
is not related to the observed flux density variability.

In an alternative interpretation we argue that the observed flux density trough can be explained
by the sum of the flux densities of the variable VLBI components. In Figure \ref{fig:lc} we display
the sum of the flux density of the modelfit components C1+C2+C3 from the VLBI observations at 15\,GHz and 22\,GHz.
Particularly at 15\,GHz, were the time coverage of the VLBI experiments covers the full time range
of the trough, there is an excellent agreement between the single dish flux density measurements and summed
VLBI component fluxes. Examining the flux densities of the individual
components (see Fig. \ref{fig:flux_year}), it appears as if the combination of the declining trend of C1
and the increasing trend of C2 are mainly responsible for the shape of the flux density trough. We therefore
conclude that the peculiar shape of the total flux density lightcurve of B\,2005+403 is a result of
a blending of the independently varying flux of the inner jet components.

\subsection{Short time-scale flux density variations}

\begin{figure}
\includegraphics[clip=,width=0.9\columnwidth]{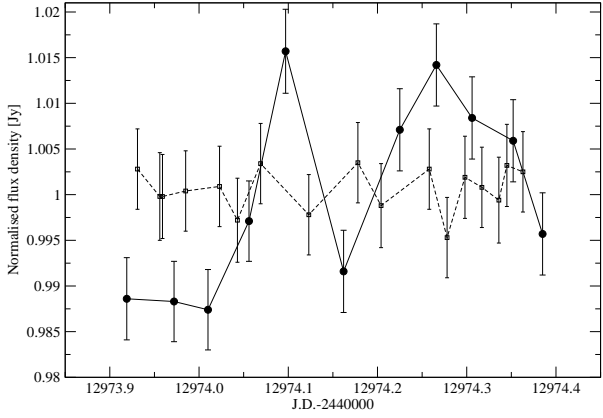}
\caption{Light-curve of B\,2005+403 at 1.6\,GHz in November 2003 observed with the
Effelsberg 100 meter telescope. Filled circles denote B\,2005+403, open squares (connected with dashed line) 
the calibrators, NGC\,7027 and B\,2021+614. The modulation indices of the secondary calibrators are a
measure of the calibration uncertainty, which is $m_0 \leq 0.25$\,\%.
The modulation index of B\,2005+403 is $m=1.01$\,\%.}
\label{fig:idv_18}
\end{figure}

\begin{figure}
\includegraphics[clip=,width=0.9\columnwidth]{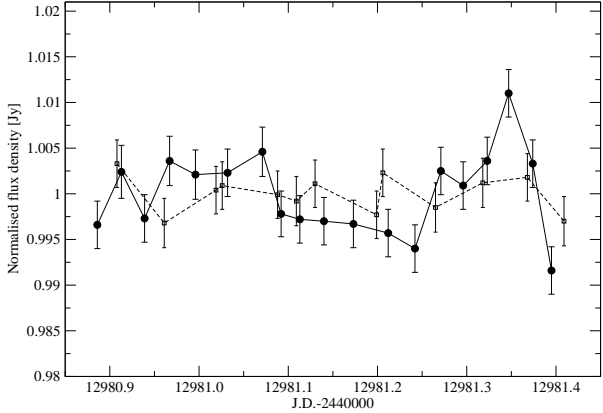}
\caption{Light-curve of B\,2005+403 at 5\,GHz in December 2003 observed with the
Effelsberg 100 meter telescope. Filled circles denote B\,2005+403, open squares (connected with dashed line) 
a calibrator, NGC\,7027. The modulation index of the calibrator is $m_0=0.20$\,\%, the modulation index
of B\,2005+403 is $m=0.45$\,\%.}
\label{fig:idv_6}
\end{figure}

In November and December 2003 the flux density variability of B\,2005+403 was monitored 
with the Effelsberg 100 meter radio telescope with high time resolution. In Fig. \ref{fig:idv_18}
and Fig.  \ref{fig:idv_6} we plot the variability curves at 18\,cm and 6\,cm, respectively.

In Table  \ref{tab:m_y} we summarize the results from these measurements. To characterize the
variability amplitudes and their significance we follow the methods described in \cite{idvh} and \cite{reduc_idv}.
Column 2 of the table gives the average flux density, col. 3 the rms standard deviation,  col. 4 the modulation
index, col. 5 the variability amplitude \citep[defined as  $Y = 3 \sqrt{m^2-m_{0}^2}$, see][]{reduc_idv}
and col. 6 the reduced $\chi^2$. A value of Y is only printed, if the $\chi^2-$test gives a higher than 99.9\,%
probability for significant variations. $m_0$ represents the maximum modulation index of the calibrators. 
From table \ref{tab:m_y} and Figures \ref{fig:idv_18} and \ref{fig:idv_6} it is obvious that 
the variability amplitude of B\,2005+403 decreases with frequency. Formally we measure a modulation index of $m=1.01$\,\%
at 1.67\,GHz and of $m=0.45$\,\% at 5\,GHz. At this frequency the detection of IDV is marginal. We note that
the model of refractive interstellar scintillation in the weak regime predicts a decrease of the modulation index
with increasing frequency \citep[e.g.][]{rick_uj}. This is consistent with our observations.

\begin{table}
\centering
\caption{\label{tab:m_y}The variability parameters of B\,2005+403 and the secondary calibrators
at 1.67\,GHz (top) and 4.85\,GHz (bottom). 
Col. 2 shows the mean flux density in Jy, col. 3 its standard deviation, col. 4 the 
modulation index, col. 5 the noise-bias corrected variability amplitude and col. 6 the reduced $\chi^2$.}
\begin{tabular}{|*6{>{$}c<{$}|}}
\hline
\text{Source name} & \langle S\rangle \text{\,(Jy)} &  \sigma  & \text{m (\%)} & \text{Y (\%)} & \text{red.} \chi^2\\ 
\hline
\multicolumn{6}{|c|}{\(\lambda=18\)\,cm, \(m_0=0.25\)\,\%} \\
\hline
\text{B\,2005+403} & 2.430 & 0.024 & 1.01 & 2.93 & 5.593 \\
\text{NGC\,7027} & 1.906 & 0.004 & 0.19 & 0 & 0.197 \\
\text{B\,2021+614} & 2.193 & 0.006 & 0.25 & 0 & 0.365 \\
\hline
\multicolumn{6}{|c|}{\(\lambda=6\)\,cm, \(m_0=0.20\)\,\%}\\
\hline 
\text{B\,2005+403} & 2.905 & 0.013 & 0.45 & 1.18 & 3.107 \\
\text{NGC\,7027} & 5.489 & 0.011 & 0.20 & 0 & 0.621 \\
\hline
\end{tabular}
\end{table}

The characteristic variability timescales derived from the light curves in Figures \ref{fig:idv_18} and \ref{fig:idv_6}
range between $t_\text{var}=0.09 - 0.13$\,days at 18\,cm and  $t_\text{var}=0.05 - 0.12$\,days at 6\,cm. Adopting
a source intrinsic interpretation for the variability we may apply the light travel time argument and
derive -- via the source size -- an apparent brightness temperature. Following \cite{ww2} we obtain for
the apparent ``variability brightness temperature'':
\[T_\text{B}=4.5\cdot 10^{10} \Delta S \left( \frac{\lambda D_\text{L}}{t_\text{var} \left(1+z\right)^2} \right)^2 \text{K}\] 
where $\Delta S$ is the amplitude of the flux density variations 
[Jy], $\lambda$ [cm] the wavelength, $D_\text{L}$ [Mpc] the luminosity distance, and $t_\text{var}$ [day] is the variability time-scale.
The resulting apparent brightness temperatures is in the range of $T_\text{B}=10^{18...21}$\,K.
This can be reduced to the inverse-Compton limit with Doppler-factors in the range of a few hundred of up to one thousand.
Since these values appear unreasonably high and are not consistent with the results from VLBI nor with theoretical
expectations \citep[e.g.][]{begelman}, we
probably can rule out that source intrinsic effects are responsible for the observed rapid variations.

The source size and the apparent brightness temperature can be also determined 
from the interstellar scintillation model, assuming that the main reason for the observed
variability is the motion of the Earth through the scintillation pattern
\citep[see for example][and references therein]{goodman}. In this scenario the variability
timescale (in [days]) is given by:
$t_\text{var} = 347.1 \theta_\text{scat} L \varv^{-1}$
with the scattering size $\theta_\text{scat}$ in [mas], the screen distance $L$ in [kpc] and the relative
velocity between screen and earth $\varv$ in [km/s]. It is clear that a screen located at kpc distance,
as proposed by \cite{gal_model} is not able to explain the observed
short variability timescale. With a scattering size (at 1.67\,GHz) of $ \theta_\text{scat} = 30$\,mas, a screen distance
of 2.4\,kpc and typical galactic velocities $\varv \leq 220$\,km/s, one would expect to see variations
on timescales $t_\text{var} > 100$ days. 

In order to reproduce a variability time scale of 0.1\,day at 1.67\,GHz we obtain with the above equation
the following constraint for scattering size and screen distance: $\theta_\text{scat} L \leq 6.34 \cdot 10^{-2}$.
For the measured scattering size of $\sim 30$\,mas, this leads to an unreasonable nearby screen distances of $\leq 2$\,pc.
The only way out of this dilemma is a smaller scattering size. Adopting a minimum screen distance
of at least 10-20\,pc as required for the interpretation of the ultrafast scintillators \citep[e.g.][]{dt2,dt3}, 
we would obtain a scattering size of $\theta_\text{scat} \leq 6.3$\,mas. A source with such a size would have an apparent brightness temperature
at 1.6\,GHz of $T_\text{B} \geq 10^{10}$\,K, fully consistent with our VLBI measurements (see Section 3.4).

An upper limit to the distance of the screen is obtained from the restriction of the source size
via the inverse Compton limit of $T_\text{B}^\text{IC}=10^{12}$\,K. The requirement that the brightness temperature has to be lower than 
this, leads to a source size of 
$\theta = \sqrt{1.22 \cdot 10^{12} S_{\nu} \nu^{-2} T_\text{B}^{-1} \delta^{-1} (1+z)} \geq 0.5$\,mas,
where we adopted $S_{\rm 1.6\,GHz} = 2.4$\,Jy from Table \ref{tab:m_y} and for the Doppler factor
$\delta \simeq 10$. This leads with the above equations to 
an upper limit of the screen distance of $D \leq  0.58 \varv$\,pc or with $(10 \leq \varv \leq 220)$\,km/s
$(6 \leq L \leq 127)$\,pc.

We therefore conclude that the observed IDV can be explained with a screen, which is located at a distance
of $L \leq 130$\,pc and which can be characterized by a scattering size of $0.5 \leq \theta_\text{scat} \leq 6.3$\,mas. 
The corresponding scattering measure at 1.6\,GHz then is of order $6.4 \cdot 10^{-4} \leq SM \leq 4.3 \cdot 10^{-2}$, measured
in the conventional units of SM. 
These values are considerably lower than the scattering measure of the more distant screen, which is responsible for
the scatter broadening in B\,2005+403. We therefore conclude that the observed IDV is not caused by this distant screen.
Most likely one has to deal with the situation of multiple scattering by at least two spatially and 
physically very different plasma screens. In this scenario the first screen leads to a significant scatter broadening 
of the source image, which in the second screen is only very weakly scattered due to large quenching effects ($\theta_\text{source}
>> \theta_\text{scat}$). This quenched scattering (e.g. \cite{rick_regi}) by the second screen could explain the
relatively low variability amplitudes of $\leq 1$\,\% observed in our IDV experiments at 6 and 18\,cm. 
Quantitatively this can be verified using equation (20) of Goodman (1997), which relates the variablity index, the
scattering measure, the effective source size and the screen distance. With the parameters from above
a modulation index of $m \leq 1$\,\% are obtained, in good agreement with our observations.

\section{Discussion and summary}

B\,2005+403 is located in the Cygnus-region and previously was known as one of the most scattered broadened
extragalactic sources in this region of the sky. In this paper we combined multi-frequency flux density monitoring and
VLBI imaging observations. We also used high time resolution flux density monitoring observations,
in order to search for possible Intra-Day Variability. The combination of these complementary data yields
new and more accurate than previously published estimates of (i) the source structure and its kinematics
and of (ii) the properties of the interstellar medium, which is responsible for the observed propagation
effects. 

High frequency VLBI imaging observations (at 15 - 43 GHz) spanning 11 years (1992 - 2003) reveal a one sided and 
bent core jet structure, with at least 5 embedded VLBI components. The inner jet components separate from the stationary 
assumed core C1 with apparent superluminal speeds of 6 - 17 c. The component nearest to the core (C2), however,
appeared stationary ($\beta_\text{app} \leq 0.3$). It is remarkable that the flux density variation of this stationary 
and steep spectrum component apparently violates the inverse Compton limit, indicating strong Doppler-boosting. 
This could be explained with a component path, which is oriented at a smaller angle to the line of sight than those 
of the other jet components. A systematic increase of the component speeds with increasing core separation is observed. 
The paths of the inner-most jet components is curved and their motion is not ballistical, suggesting a spatially
curved (helical) jet. Beyond 2\,mas core-separation, however, the components seem to move on linear (ballistical) 
trajectories. Continued future VLBI monitoring will be necessary to determine their paths more accurately.

The VLBI images at longer wavelengths (1.6 - 8 GHz) show a scatter broadened nearly point-like source.
Combining our new size measurements with published data from earlier observations allowed to study the frequency dependence
of the source size in greater detail. Below 8 GHz this dependence is best described by a non-Kolmogorov 
power law with slope of -1.9. For the size of the scattering disk at 1\,GHz we obtain 
$\left( 77.1 \pm 4.0 \right) $\,mas.
Above 8\,GHz the measured sizes are increasingly larger than the prediction from the
scattering law, and the internal source structure becomes visible.
Based on the observed scatter broadening, several parameters of the scattering medium (scattering size, scattering
measure, electron density) were determined and found to be in accordance with previously published 
estimates (NE2001 model, Cordes \& Lazio, 2002). The NE2001 model places the scattering screen
at a distance of 2.35\,kpc. We however note that towards the direction of  B\,2005+403 there is a
degree-size extended supernova remnant \object{G78.2+2.1}. It
exhibits a patchy, inhomogeneous structure at optical, X-ray and radio wavelengths \citep[][and references therein]{snr_xray}.
Its distance is $ \left( 1.5\pm0.5 \right)$ \,kpc \citep{snr_dist}, still compatible with the scattering model.

We discuss the observed long term flux density variability curves (5 - 37\,GHz), which during 1996 - 2001 showed a remarkable
flux density 'trough'. A tentative interpretation via a possible 'scattering event', similar to the 
'extreme scattering events' observed in other sources, could be abandoned on the basis of the observed
frequency dependence of the variability, which appeared more pronounced at higher frequencies. 
Instead, we show that the 'trough' is well explained by the
summed flux density variability of the (independently) varying VLBI components (C1 - C3). In contrast
to other sources, which often show correlations between jet component ejection and flux density variability
\citep[e.g.][and references therein]{var},
such a relation is not seen B\,2005+403. In this source at least some (if not most) of the observed 
flux density variability results from the blending of the evolving jet components.

Dense in time sampled variability measurements performed with the 100\,m Effelsberg telescope
showed only weak intra-day variability in B\,2005+403. This was in contrast to naive expectations, in which
more pronounced variability was anticipated. The low variability indices ($m=1$\,\% at 18\,cm, $m=0.5$\,\%
at 6\,cm) and the short variability time scale of $\sim 0.1$\,days cannot be explained by scattering effects
from the kpc-screen, which is responsible for the scatter broadening. Instead, another and much closer scatterer
is required. Using the thin screen approximation for refractive interstellar scintillation we determined
the likely distance of the scattering screen to be $(6 \leq L \leq 128)$\,pc. At 18\,cm, its scattering 
size would lie in the range of  $0.5 \leq \theta_\text{scat} \leq 6.3$\,mas. For B\,2005+403 one therefore has
to deal with the situation that a scattered broadened image (kpc-screen) is scattered a second time (pc-screen) and 
that the combination of both effects limits the accuracy of low frequency flux density variability measurements 
to about 0.5-1\,\% and the angular the resolution of VLBI observations to a few milliarcseconds. We note that weak
refractive interstellar scintillation from a very nearby and ionized ISM (which most likely covers larger portions of 
the sky), may therefore also limit the calibration accuracy and sensitivity of planned large radio telescopes
operating and low frequencies, like LOFAR or the square-kilometer array (SKA).

\begin{acknowledgements}
We wish to thank M. Lister, and Hugh Aller and Margo Aller for kindly providing data prior to publication.
We like to thank Alan Roy, Thomas Beckert and Jacques Roland for scientific discussion and useful comments.
This work is based on observations with the 100 meter telescope of the MPIfR 
(Max-Planck-Institut f\"ur Radioastronomie) at Effelsberg.
This research has made use of data from the University of Michigan Radio Astronomy Observatory 
which has been supported by the University of Michigan and the National Science Foundation. The European VLBI Network 
is a joint facility of European, Chinese, South African and other radio astronomy institutes funded by 
their national research councils. The National Radio Astronomy Observatory is a facility of the 
National Science Foundation operated under cooperative agreement by Associated Universities, Inc.
U.B. was partly supported by the European Community's Human Potential Programme under contract HPRCN-CT-2002-00321.
K.\,\'E.\,G. has been supported for this research through a stipend from the International 
Max Planck Research School (IMPRS) for Radio and Infrared Astronomy at the Universities of Bonn and Cologne.
\end{acknowledgements}

\bibliographystyle{aa}
\bibliography{irod}

\Online
\appendix

\setcounter{table}{0}

\begin{longtable}{c|c*6{>{$}c<{$}}|c}

\caption[]{\label{tab:model15}The parameters of the Gaussian modelfits. 
Column 1 gives the observing epoch, the VLBI array and the frequency, column
2 the component identification. Column 3 shows the flux density of each component. In Col. 4, the relative separation
to C1 and in Col. 5 the position angle are listed. Columns 6, 7 and 8 list the 
major axis, the axial ratio of minor to major axis, and the major axis position angle. The last column 
shows the reduced $\chi^2$ of the model.} \\
Epoch, array and frequency & Comp. & S\ \mbox{(Jy)} & r\  \mbox{(mas)} &  \theta\  (^\circ) & a\ \mbox{(mas)} & b/a & \phi\  (^\circ) & $\chi^2_r$ \\ 
(1) & (2) & (3) & (4) & (5) & (6) & (7) & (8) & (9) \\
\hline \hline
\endfirsthead

\caption[]{continued} \\
Epoch, array, frequency & Comp. & S\ \mbox{(Jy)} & r\  \mbox{(mas)} &  \theta\  (^\circ) & a\ \mbox{(mas)} & b/a & \phi\  (^\circ) & $\chi^2_r$ \\ 
(1) & (2) & (3) & (4) & (5) & (6) & (7) & (8) & (9) \\
\hline \hline
\endhead
1998.14 & M1 & 2.365 & 0.00 & - & 29.76 & 0.8 & 37.5 & 4.5 \\
EVN & & & & & & & & \\
1.6\,GHz & & & & & & & & \\
Errors & & 4\,\% & & & 7\,\% & 3\,\% & 3^\circ & \\
\hline
\hline
1996.82 & M1 & 2.572 & 0.00 & - & 3.48 & 0.70 & 32.1 & 2.6 \\
EVN & M2 & 0.488 & 2.67 & 129.7 & 3.24 & 1.00 & - & \\
5\,GHz & M3 & 0.069 & 8.10 & 138.3 & 6.95 & 1.00 & - & \\
Errors & & 10\,\% & 12\,\%& 4^\circ & 5\,\% & 10\,\% & 4^\circ \\
\hline\hline
1996.83 & M1 & 3.418 & 0.00 & - & 1.51 & 0.86 & 65.43 & 1.0 \\
EVN & M2 & 0.297 & 3.06 & 127.0 & 2.13 & 0.69 & 44.7 & \\
8\,GHz & & & & & & & \\
Errors & & 11\,\% & 6\,\% & 3^\circ & 6\,\% & 10\,\% & 5^\circ & \\
\hline
\hline
1995.27 & C1 & 0.852 & 0.00 & - & 0.40 & 1.00 & - & 1.0 \\
VLBA    & C2 & 0.755 & 0.39 & 90.0 & 0.37 & 1.00 & - & \\
15\,GHz & C3 & 0.666 & 0.77 & 74.8 & 0.37 & 1.00 & - & \\
        & C4 & 0.170 & 1.78 & 122.0 & 1.44 & 1.00 & - & \\
        & C5 & 0.185 & 3.35 & 116.7 & 1.79 & 1.00 & - & \\
Errors  & & 10\,\% & 14\,\% & 3^\circ & 12\,\% & & & \\
\hline
1995.96 & C1 & 0.510 & 0.00 & - & 0.31 & 1.00 & - & 0.3 \\
VLBA    & C2 & 0.831 & 0.38 & 97.0 & 0.35 & 1.00 & - & \\
15\,GHz & C3 & 0.951 & 0.84 & 80.4 & 0.40 & 1.00 & - & \\
        & C4 & 0.106 & 2.12 & 118.9 & 1.34 & 1.00 & - & \\
        & C5 & 0.104 & 3.72 & 116.6 & 1.55 & 1.00 & - & \\
Errors  & & 10\,\% & 12\,\% & 4^\circ & 10\,\% & & & \\
\hline 
1996.73 & C1 & 0.260 & 0.00 & - & 0.28 & 1.00 & - & 0.3 \\
VLBA+EB & C2 & 0.738 & 0.40 & 108.5 & 0.44 & 1.00 & - & \\
15\,GHz & C3 & 0.981 & 0.92 & 83.1 & 0.45 & 1.00 & - & \\
        & C5 & 0.185 & 3.16 & 119.0 & 2.38 & 1.00 & - & \\
Errors  & & 10\,\% & 18\,\% & 3^\circ & 10\,\% & & & \\
\hline
1997.19 & C1 & 0.262 & 0.00 & - & 0.25 & 1.00 & - & 2.0 \\
VLBA    & C2 & 0.659 & 0.42 & 107.5 & 0.47 & 1.00 & - & \\
15\,GHz & C3 & 0.791 & 0.99 & 84.1 & 0.48 & 1.00 & - & \\
        & C4 & 0.075 & 2.38 & 119.0 & 1.44 & 1.00 & - & \\
        & C5 & 0.092 & 4.07 & 117.8 & 1.73 & 1.00 & - & \\
Errors  & & 15\,\% & 11\,\% & 3^\circ & 7\,\% & & & \\
\hline
1999.01 & C1 & 0.396 & 0.00 & - & 0.38 & 1.00 & - & 0.5 \\
VLBA    & C2 & 0.510 & 0.61 & 101.4 & 0.68 & 1.00 & - & \\
15\,GHz & C3 & 0.422 & 1.18 & 80.4 & 0.67 & 1.00 & - & \\
        & C4 & 0.234 & 2.76 & 117.9 & 3.62 & 1.00 & - & \\
Errors  & & 12\,\% & 17\,\% & 3^\circ & 3\,\% & & \\
\hline
2001.17 & C1 & 0.024 & 0.00 & - & 0.17 & 1.00 & - & 3.5 \\
VLBA    & C2 & 1.800 & 0.35 & 138.3 & 0.39 & 0.80 & 38.2 & \\
15\,GHz & C3 & 0.764 & 1.31 & 94.3 & 0.83 & 0.69 & 47.0 & \\
        & C5 & 0.179 & 4.21 & 121.3 & 4.14  & 1.00 & - & \\
Errors  & & 7\,\% & 15\,\% & 3^\circ & 3\,\% & 4\,\% & 5^\circ \\
\hline
2001.98 & C1 & 0.091 & 0.00 & - & 0.20 & 1.00 & - & 1.6 \\
VLBA    & C2 & 1.581 & 0.37 & 137.9 & 0.45 & 0.70 & 36.0 & \\
15\,GHz & C3 & 0.632 & 1.42 & 102.9 & 0.77 & 0.74 & 27.2 & \\
        & C4 & 0.172 & 3.06 & 112.7 & 3.66 & 1.00 & - & \\
Errors  & & 5\,\% & 10\,\% & 3^\circ & 6\,\% & 4\,\% & 4^\circ \\
\hline
2003.04 & C1 & 0.188 & 0.00 & - & 0.2 & 1.0 & - & 0.3 \\
VLBA    & C2 & 1.714 & 0.37 & 118.0 & 0.40 & 0.79 & 24.2 & \\
15\,GHz & C3 & 0.516 & 1.56 & 97.9 & 0.86 & 0.81 & 29.0 & \\
        & C4 & 0.008 & 2.63 & 131.4 & 0.27 & 1.00 & - & \\
        & C5 & 0.061 & 4.59 & 121.5 & 2.23 & 1.00 & - & \\
Errors  & & 6\,\% & 15\,\% & 3^\circ & 6\,\% & 4\,\% & 4^\circ \\
\hline
2003.16 & C1 & 0.223 & 0.00 & - & 0.34 & 0.45 & 30.3 & 1.5 \\
VLBA    & C2 & 2.000 & 0.41 & 120.6 & 0.41 & 0.79 & 25.1 & \\
15\,GHz & C3 & 0.566 & 1.60 & 99.5 & 0.85 & 0.89 & 40.7 & \\
        & C4 & 0.017 & 3.00 & 124.4 & 0.99 & 1.00 & - & \\
        & C5 & 0.034 & 4.84 & 122.0 & 1.63 & 1.00 & - & \\
Errors  & & 4\,\% & 12\,\% & 3^\circ & 4\,\% & 5\,\% & 4^\circ \\
\hline
\hline
1992.44 & C1 & 0.337 & 0.00 & - & 0.21 & 0.63 & 88.3 & 2.5 \\
EVN & C2 & 0.723 & 0.35 & 97.4 & 0.43 & 0.79 & 37.4 & \\
22\,GHz  & C3 & 0.695 & 0.65 & 87.7 & 0.20 & 0.77 & 71.2 & \\
 & C5 & 0.217 & 2.37 & 117.3 & 2.29 & 0.10 & 86.0 & \\
Errors & & 10\,\% & 10\,\% & 3^\circ & 10\,\% & & 5^\circ & \\
\hline
1994.17 & C1 & 0.570 & 0.00 & - & 0.27 & 0.68 & -54.3 & 6.3 \\
VLBA+ & C2 & 0.796 & 0.40 & 87.7 & 0.28 & 0.56 & 62.8 & \\
+VLA+EVN & C3 & 0.526 & 0.81 & 82.4 & 0.24 & 0.85 & 22.7 & \\
22\,GHz & C5 & 0.157 & 2.56 & 112.7 & 2.32 & 0.92 & -2.7 & \\
Errors & & 18\,\% & 12\,\% & 3^\circ & 18\,\% & & 4^\circ & \\
\hline
1996.73 & C1 & 0.434 & 0.00 & - & 0.29 & 0.65 & -53.1 & 4.1 \\
VLBA+EB & C2 & 0.431 & 0.42 & 102.9 & 0.27 & 0.69 & 33.1 & \\
22\,GHz & C3 & 0.667 & 0.85 & 80.9 & 0.35 & 0.91 & 53.1 & \\
        & C4 & 0.212 & 2.12 & 125.7 & 4.53 & 0.28 & -72.4 & \\
Errors  & & 6\,\% & 11\,\% & 3^\circ & 18\,\% & & 5^\circ \\
\hline 
2003.04 & C1 & 0.329 & 0.00 & - & 0.20 & 1.0 & - & 0.9 \\
VLBA    & C2 & 1.339 & 0.37 & 121.0 & 0.21 & 0.77 & 27.2 & \\
22\,GHz & C3 & 0.400 & 1.52 & 98.1 & 0.78 & 0.89 & 28.2 & \\
        & C4 & 0.015 & 2.77 & 127.1 & 0.41 & 1.0 & - & \\
Errors  & & 6\,\% & 15\,\% & 3^\circ &  8\,\% & & 3^\circ \\
\hline
\hline
1996.73 & C1 & 0.343 & 0.00 & - & 0.30 & 0.57 & -36.3 & 0.6 \\
VLBA+EB & C2 & 0.304 & 0.46 & 102.6 & 0.23 & 0.67 & -0.2 & \\
43\,GHz  & C3 & 0.442 & 0.92 & 83.0 & 0.33 & 0.84 & 36.2 & \\
 & C5 & 0.191 & 3.73 & 116.4 & 3.01 & 0.66 & 35.5 & \\
Errors & & 7\,\% & 6\,\% & 3^\circ & 12\,\% & & 5^\circ & \\
\hline
\end{longtable}

\end{document}